# Impact of intergrain spin transfer torques due to huge thermal gradients on the performance of heat assisted magnetic recording

Bernard Dieny[1a], Mair Chshiev[1], Brian Charles[1], Nikita Strelkov[1,4], Alain Truong[1], Olivier Fruchart[1], Ali Hallal[1], Jian Wang[2], Yukiko K. Takahashi[2], Tomohito Mizuno[3], Kazuhiro Hono[2]

1. Univ. Grenoble Alpes, CEA, CNRS, Grenoble INP*, INAC-SPINTEC, 38000 Grenoble, France

*Institute of Engineering Univ.Grenoble Alpes

2. National Institute for Materials Science (NIMS), 1-2-1 Sengen, Tsukuba 305-0047, Japan

3. TDK/Headway, 682 South Hillview Drive, Milpitas, CA 95035

4. Department of Physics, Lomonosov Moscow State University, Moscow 119991, Russia

a. Email: bernard.dieny@cea.fr

**Abstract**

Heat assisted magnetic recording (HAMR) is a new technology which uses temporary near field laser heating of the media during write to increase hard disk drive storage density. By using plasmonic antenna embedded in the write head, extremely high thermal gradient are created in the recording media (up to 10K/nm). State of the art HAMR media consists of grains of FePtX ordered alloys exhibiting high perpendicular anisotropy separated by grain boundaries, typically made of a segregant such as $SiO_2$, $SiN_x$ or carbon 1 to 2nm thick. Nearby the plasmonic antenna, the difference of temperature between two 8nm wide neighboring grains in the media can reach 80K across the 2nm thick grain boundary, representing a thermal gradient of 40K/nm. This represents a gigantic thermal gradient across a tunnel barrier. Such situations with much weaker thermal gradient (~1K/nm, already considered as very large) have already been studied in the field of spincaloritronics. There, it was shown that very large spin transfer torques due to thermal gradients can arise in magnetic tunnel junctions, larger than spin transfer torque due to electrical currents. These torques can even yield magnetization switching in magnetic tunnel junctions. Considering that two neighboring grains separated by an insulating grain boundary in a HAMR media can be viewed as a magnetic tunnel junction and that the thermal gradients in HAMR are one to two orders of magnitude larger than those existing in conventional spincaloritronics, one may expect a major impact from these thermal torques on magnetization switching dynamics and therefore on HAMR recording performances. This issue has never been considered so far in the development of HAMR technology. This paper combines theory, experiments aiming at determining the polarization of tunneling electrons across the media grain boundaries, and micromagnetic simulations of recording process taking into account these thermal gradients. It is shown that the thermal in-plane torque can have a detrimental impact on the recording performances by favoring antiparallel magnetic alignment between neighboring grains during the media cooling. Implications on media design are discussed in order to limit the impact of these thermal torques.

**Introduction to heat assisted magnetic recording**

Over the past 60 years, a tremendous increase by almost 10 orders of magnitude in magnetic areal density of storage on hard disk drives as well as an associated decrease of cost per Gigabyte were achieved[1]. Together with FLASH memory based Solid State Drives, these complementary technologies allowed to face the exponentially increasing quantity of data that our society generates and requires to store. Nowadays, the prevailing recording technology is the perpendicular magnetic recording (PMR). It uses perpendicular media, i.e. granular media having large out-of-plane anisotropy. This technology is approaching areal density of 1Terabit/in² (i.e. 1000Gbit/in²)[1]. However, it is gradually reaching a physical limit due to the so called magnetic recording trilemma[2]. This trilemma can be understood as follows: Perpendicular recording media consist of an assembly of coercive nanometer-sized grains of average volume V, separated by grain boundaries. The grain boundaries are made of a segregant (e.g.Ta) aiming at magnetically decoupling the grains from each other. Typical grain size diameter in PMR media is of the order of 6nm for a thickness of about 10nm [3]. The magnetic grains are made of CoCrPt-based alloys which have an hcp structure with out-of-plane c-axis conferring them an out-of-plane magnetic anisotropy per unit volume K. The stability of the written information over 10 years requires the energy barrier separating the two stable states of magnetization in a grain to be large enough against thermal activation energy. This condition writes $KV>40k_BT$ where $k_B$ is the Boltzmann constant and T the operation temperature. Writing a bit (0 or 1) in the media consists in setting the magnetization of a small set of neighboring grains in the up or down direction thanks to the magnetic field delivered by the poles of the write head. The number *N* of grains per bit has significantly decreased from ~100 to ~10 over the years thanks to improvements in the readout electronic channel[2]. However this number cannot be decreased much more since the media signal to noise ratio in these granular media is given by $SNR \sim 10\log_{10}(N)$. Increasing the areal storage density in hard disk drives implies reducing the bit size and therefore reducing the grain size $V$. However, maintaining the thermal stability of the grain then requires to increase the media anisotropy. This then yields a problem of writeability. Indeed, in macrospin approximation, the field required to switch the magnetization of a grain is given by $\mu_0 H_0 = \frac{2K}{M_s} - \mu_0 N_{eff} M_s$ where $\mu_0$ is the vacuum permeability, $N_{eff}$ the grain demagnetizing coefficient and $M_s$ the grain magnetization (SI units). Increasing the grain anisotropy thus requires to produce larger magnetic field to switch the grain magnetization upon writing. However, a write head cannot produce fields larger than about 2.4T because this field is essentially limited by the saturation magnetization of the head pole pieces and no material can deliver a field larger than ~2.4T. The recording trilemma therefore results from the combination of three requirements: i) the media SNR imposing a minimum number of grains per bit, ii) the thermal stability of the grains requiring to increase their anisotropy as the grain size is reduced, iii) the writeability necessitating to maintain the write field below the maximum field that the head can provide[2]. To circumvent this trilemma, various approaches have been proposed[1]. They can be classified in three categories[4]. The first category consists in upgrading the PMR media, for instance by using exchange coupled composite media consisting of coupled hard and soft layers[5,6]. Thanks to an easier nucleation of the magnetization reversal in the soft part of the grains while maintaining the energy barrier, this allows to lower the magnetization switching field for a given grain thermal stability. The use of a soft underlayer (SUL) below the recording layer allows also to concentrate the magnetic flux towards the written bit and increase the head write efficiency[7]. These features are actually implemented in commercial PMR hard disk drives. A second category of approaches consists in using bit patterned media (BPM) instead of granular media. The media is then nanostructured in the form of a regular array of nanometer size single grain magnetic dots, each dot carrying one bit of information[8,9,10,11,12]. In laboratory demonstrations, the technology has proven to work quite well and would allow reaching areal density of several Tbits/in². However, for patterned media to become an economically viable alternative, means to fabricate large area patterns of such high density at reasonable costs have to be developed. Nano imprint lithography combined with diblock copolymers is the preferred investigated approach. Still, the economic viability of BPM in regards to competing storage technology, in particular FLASH based storage, remains uncertain. The third category of approaches consists in assisting the magnetization switching during write by various means. Two main routes are being investigated. The first one is based on microwave assistance. The idea is to provide to the grain magnetization additional energy on top of the Zeeman energy brought by the write field, to help the magnetization overcome the energy barrier which separates its two stable states[13,14,15]. In this technology, the microwave is produced by a spin transfer oscillator embedded in the gap of the write head[13,14]. Great progress has been made lately on this technology. Western Digital recently announced the commercialization of MAMR based hard disk drives[16]. Another energy-assisted technology likely more extendable in terms of storage areal density is the heat assisted magnetic recording (HAMR)[17,18,19,20,21]. It consists in temporarily heating the media during write to lower or even suppress the grain anisotropy energy and therefore the energy barrier separating the two stable states of magnetization. Media with much higher anisotropy at room temperature than in PMR can then be used and be written at elevated temperature by the limited field provided by the write head. Since the heated area has to be of the order of the bit size (a few tens of nm), the heating is produced by a near field electromagnetic waves produced by a plasmonic antenna integrated in the gap of the write head (Figure.1). In a hard disk drive, the disk typically rotates at 7000 to 10000 revolution per minute (rpm), representing linear velocity of the order of 20m/s. As the media passes underneath the write head and approaches the plasmonic antenna, it locally heats up to a temperature close to or even slightly above its Curie temperature (400°C to 500°C) within a few ns. The anisotropy field of the heated grains then drops below the write field of the head. As the media moves away from the plasmonic antenna, it cools back to the standby temperature within a few ns. During the initial stage of the cooling after the media crosses the Curie temperature (~first 100ps), the grain magnetization starts growing along the local magnetic field. Then, as the anisotropy field very rapidly grows, the grains magnetization precesses around this anisotropy field and due to the Gilbert damping, progressively gets aligned along the vertical anisotropy axis. In HAMR, to avoid too large precession angle which may yield write error due to thermal fluctuations during the cooling process, it is preferable to design the write head so that the write field is close to vertical at the line where the media crosses its Curie temperature. This allows the magnetization to start growing below Tc along its anisotropy direction thereby minimizing its ringing during cooling and correlatively minimizing the write error rate[22].

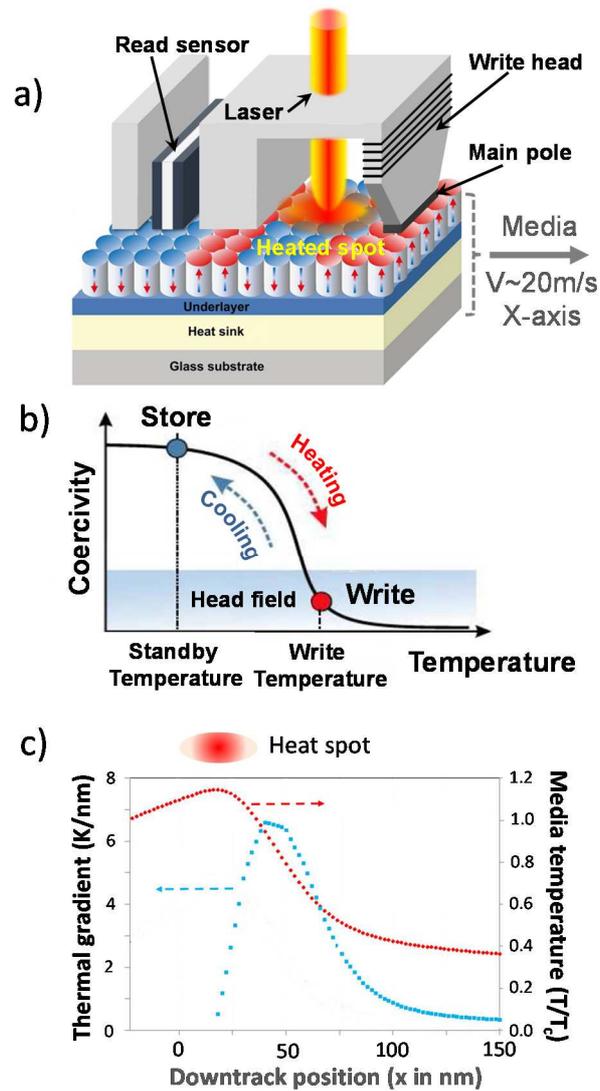

Fig.1: Basic principles of HAMR. a) Sketch of HAMR write and read head, and media; b) temperature variation of media coercivity. Storage takes place at the standby temperature (ambient temperature) while writing is performed at elevated temperature, while cooling from slightly above the media Curie temperature. c) Downtrack variation of the temperature in the recording media (right scale) normalized by the media Curie temperature and corresponding thermal gradient (left scale).

HAMR media are granular media like PMR media but with a few major differences. They are made of higher anisotropy materials at room temperature than PMR media. Typically FePtX-Y based ordered alloys are being used (X=Cu, Ag; Y=carbon or other segregants e.g. BN, $TiO_2$, $Cr_2O_3$, $SiO_2$)[3,23]. Introducing Cu or Ag in FePt allows tuning the Curie temperature of this ordered alloy from 725K for $Fe_{55}Pt_{45}$ to 600K for $Fe_{33}Cu_{13}Pt_{54}$[3]. A tradeoff must be found on Tc. It should be high enough so that the grain anisotropy energy is large enough to ensure information retention for 10 years at the disk standby temperature. However, it should not be too high to avoid degradation of the lubricant at the disk surface during write and excessive heating of the head yielding reliability issues. Typically, Curie temperatures in the range 620K to 670K are desirable. To improve recording performances, great progress were achieved to narrow down the grain to grain temperature distribution which is influenced by the size and degree of ordering of the grains[24,25,26] as well as to minimize the proportion of grains having misoriented (i.e. in-plane oriented) anisotropy axis[27,28,29]. Also, by changing the segregant material constituting the intergrain boundaries and using dual magnetic layers, the formation of cylindrical rather than spherical [22, 27, 29,30] magnetic grains could be favored to maximize the volume fraction of magnetic material in the recording layer.

From a thermal point of view, the media is designed to avoid lateral spread of the heat and in contrast favor fast dissipation towards the underlying substrate. To achieve this, the grain boundaries are preferably made of a material with low thermal conductivity and a heat sink layer is deposited underneath the recording layer to increase the downwards vertical heat flow throughout the media.

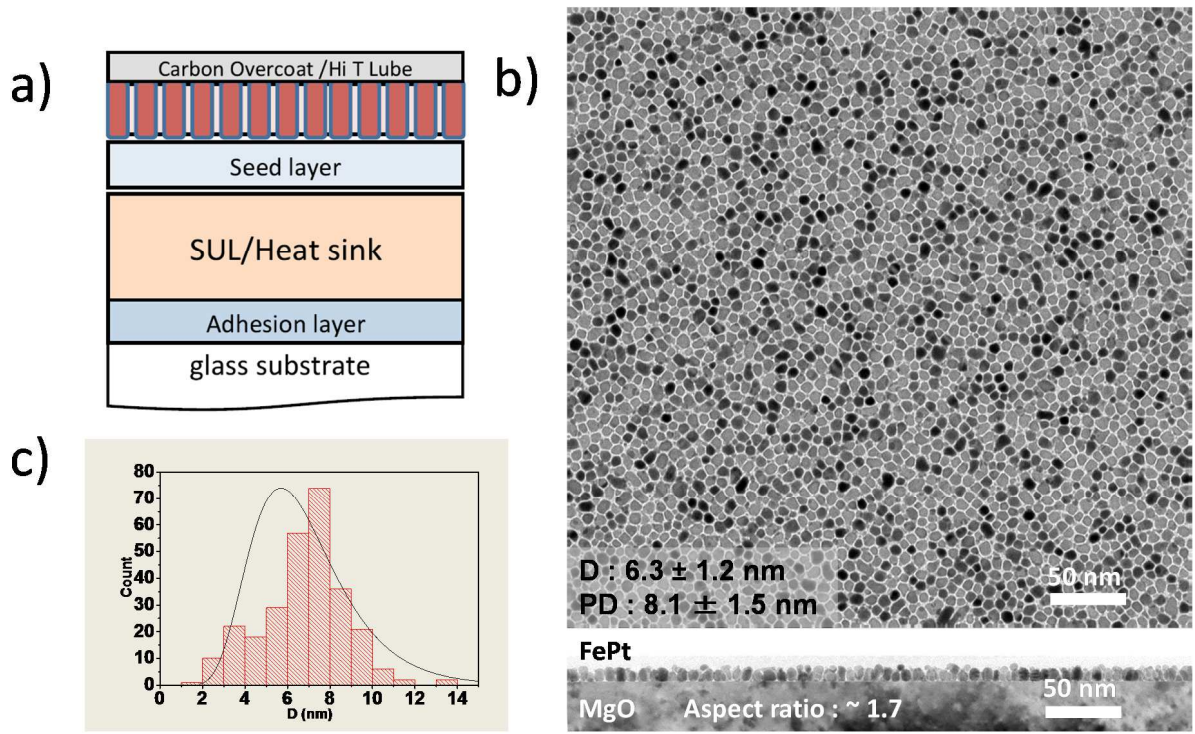

*Fig.2: a) Cross-sectional schematic representation of a HAMR media. From the glass substrate, the HAMR media comprises an adhesion layer, a heat sink and soft underlayer (SUL), a seed layer to promote the growth of the granular magnetic recording layer with perpendicular anisotropy and a carbon overcoat b) Top view and cross-sectional view of a MgO/FePtC media used in the experimental part of this study; c) distribution of grain size in the media of Fig.b).*

A typical HAMR media[3] is shown in Fig.2. On a glass substrate, it first comprises an adhesion layer (e.g.NiTa) and a heat sink layer (e.g. Ag, Au, Cr). A soft underlayer (SUL) is commonly added to enhance the vertical component of the write field on the recording layer. On top of this layer, a seed layer is deposited promoting the growth of the recording layer with large perpendicular anisotropy. With FePtX based alloys, the seed layer is intended to favor the $L1_0$ FePt(100) perpendicular orientation of the magnetic layer. Various underlayers were investigated (e.g. Pt/Cr, CrRu, RuAl, TiN, Cr, Ti, TiC and MgO) with a preference for MgO[3]. In state of the art HAMR media, the recording layer is a granular FePtCu alloy with C segregants forming the grain boundaries. The grain size is typically around 8nm with grain boundaries 1 to 2nm thick. C with possible addition of another element[3] is currently the preferred segregant because it favors the formation of a narrow distribution of well oriented FePtX grains having a close to cylindrical shape with an almost uniform thickness of the intergranular segregant material.

As explained previously, in the HAMR writing process, the information is written during the initial stage of the cooling of the grains below their Curie temperature as the media locally moves away from the heating transducer (i.e. the plasmonic antenna). In perpendicular magnetic recording, to get narrow transitions between bits, the write head is designed to generate a very strong write field gradient $\frac{dH}{dx}$ in the recording layer (x-axis is along the downtrack direction, see Fig.1). In HAMR, since $\frac{dH}{dx} = \frac{dH_c}{dT}\frac{dT}{dx}$, $H_c$ representing the media coercive field, this condition on the write field gradient is replaced by a condition on the thermal gradient. The plasmonic antenna and the media are both designed to create a very large thermal gradient in the media underneath the write head, this thermal gradient being combined with the largest possible vertical component of write field. In state of the art HAMR head and media, thermal gradient of the order of 10K/nm can be achieved combined with vertical write fields of 0.5 to 0.8 T[29]. This value of 10K/nm is a spatial average of the thermal gradient. Since the grains are metallic (high thermal conductivity) and the grain boundaries are insulating or semiconductor (comparatively lower thermal conductivity), most of the longitudinal temperature gradient takes place across the grain boundaries. Considering that the grain size is about 8nm, this means that a temperature difference of 80K can exist across a grain boundary 1 to 2 nm thick yielding thermal gradients across the grain boundaries in the range 40 to 80K/nm!

At this point, one has to realize that these thermal gradients in HAMR technology are absolutely huge. They can be expected to yield spincaloritronic effects in the media such as intergrain thermal spin transfer torques which have never been discussed so far in HAMR and can significantly impact the recording performances as explained in the following sections.

**Introduction to spin transfer torques due to thermal gradients**

The field of spincaloritronics emerged about 15 years ago. It deals with new phenomena which combine charge, spin and heat currents. Following early works on magnetothermoelectric power and Peltier effect in magnetic nanowires[31,32,33] and nanopillars[34], it greatly expanded after the prediction and observations that temperature gradients can generate spin transfer torque (called thermal spin-transfer torque: TST)[35,36,37]. Thermal spin transfer torques driven by heat current are in fact very efficient: using parameters typical of Permalloy, it was calculated that a temperature gradient of 0.2K/nm is as efficient in terms of spin transfer torque as a charge current of density $10^7$A/cm² [38]. In spincaloritronics experiments, the thermal gradients which are typically generated by Joule heating or laser lightening are typically in the range $10^{-4}$K/nm[35] in metallic nanowires and up to ~1K/nm at low temperature across the tunnel barrier of MgO-based magnetic tunnel junctions (MTJ)[39]. Such thermal gradient of 1K/nm across an oxide barrier was already considered as very large[39] while the thermal gradients in HAMR are almost two orders of magnitude larger !thermoeledtr The 1K/nm thermal gradient in MTJs was shown to produce a spin transfer torque comparable to the one due to a pure charge current of about $10^6$A/cm² [39]. This TST is sufficiently large to significantly influence the switching of the free layer magnetization in MTJ. In these experiments[39], the free layer was hotter than the reference layer which was shown to favor antiparallel magnetic configuration. We will come back to this point later in the context of HAMR writing.

Considering the huge thermal gradients produced in HAMR and the potentially very large TST which can result from these thermal gradients in the media, we investigated the impact of this effect on the HAMR recording process and performances.

The study comprised several steps which are reported in the following sections of the paper. In free electron model, we first calculated the spin transfer torque in an MTJ whose magnetic electrodes are at different temperatures. Such an MTJ is used as a model for two adjacent ferromagnetic grains at different temperatures separated by an insulating or semiconductor grain boundary. To calibrate the model to actual FePtC HAMR media, we then experimentally estimated the spin polarization of the electrons tunneling from grain to grain across C grain boundaries by measuring the current in-plane magnetoresistance of the HAMR media in combination with magnetic force microscopy. From the obtained tunneling spin-polarization, the amplitude of the thermal in-plane and perpendicular spin transfer torques between grains during the media cooling could be determined from the model. Micromagnetic simulations based on the LLG equation including the calculated thermal spin transfer torques were then performed. The impact of the TST on the recording performances was investigated by varying the TST amplitudes around the value predicted by our modelling. The simulations indicate that this impact is detrimental for the recording performance and can be quite significant. Recommendations on the media design to minimize this impact are given.

**Modeling of thermal spin transfer torque in magnetic tunnel junctions**

Two FePt neighboring grains of the HAMR media separated by an insulating barrier (e.g. carbon, 2nm) moving underneath the plasmonic antenna were assimilated to a magnetic tunnel junction in which the two magnetic electrodes are at different temperatures. The calculation of the ballistic tunnel currents and torques was performed in the free electron model with parabolic dispersion laws using non-equilibrium Green function technique, Keldysh formalism and WKB approximation[40]. The MTJ structure was assumed to have semi-infinite FM electrodes. The magnetization of one ferromagnetic electrode was fixed whereas the magnetization of the other electrode could be set at any angle with respect to the other one.

The current J(V, ΔT) through the tunnel barrier as well as the in-plane and perpendicular to the plane components of the spin transfer torque were first calculated as a function of bias voltage V and difference of temperature ΔT between the two electrodes.

The parameters of the system were chosen to get close to the real materials: Adopting the itinerant d-like electrons model first proposed by Stearns[41] and based on ab-initio calculations of the electronic band structure of FePt[42], a band splitting of 2Δ=2.2eV was chosen with the Fermi energy $\varepsilon_F$ being at $\Delta_0$=2.0eV above the center of the gap between spin-up and down bands. Correspondingly, the parameters in Fig.3a takes the value $U_L^{\downarrow} = 0.9 eV$ and $U_L^{\uparrow} = 3.1 eV$ which yields a Stearns polarization of P=0.30. The barrier height for carbon was set to $U_B$=0.14eV. By varying the barrier thickness, the resistance x Area product (RA) of the MTJ can be tuned to match the experimentally determined value for the FePtC grain boundaries, measured to be 0.30Ω.µm² as we will see in the next section.

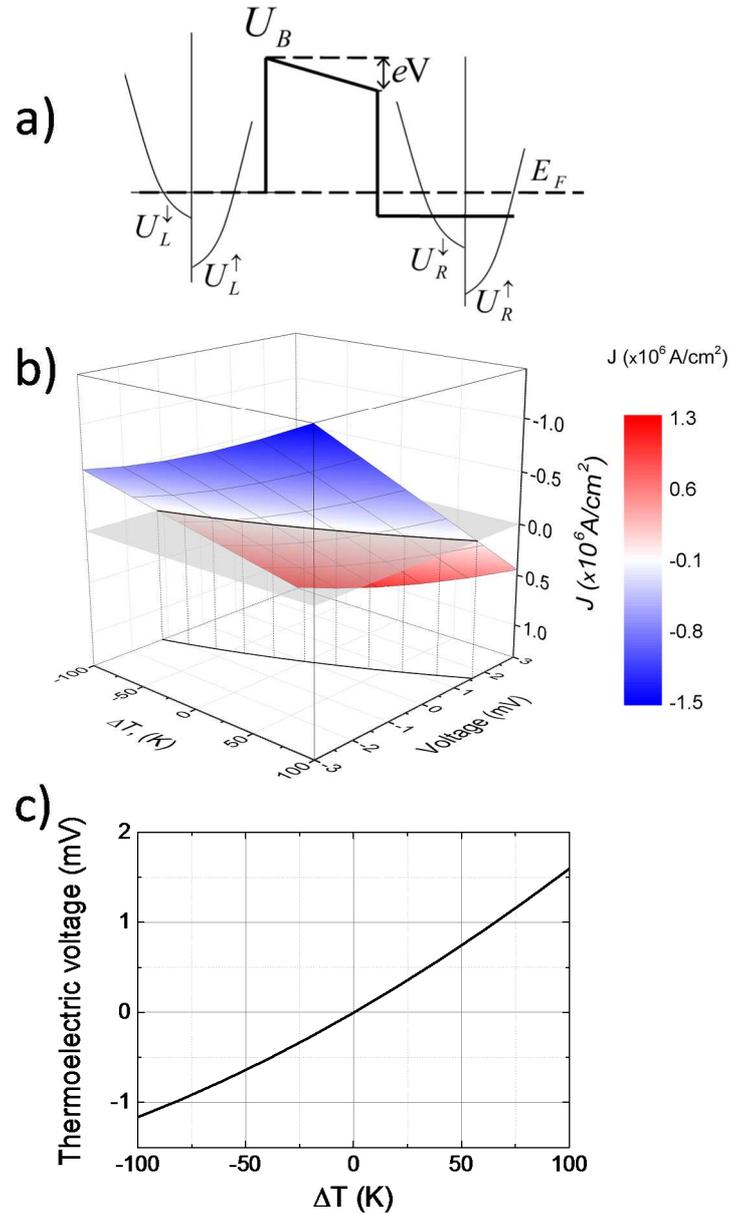

*Figure3: a) Schematic representation of the band structure (band-splitted free electron model) and of the tunnel barrier to define the parameters used in the model. b) Ballistic tunneling current across the insulating barrier in magnetic tunnel junctions in parallel magnetic configuration versus voltage and difference of temperature between the two electrodes (ΔT). c) Thermoelectric voltage across the tunnel barrier versus ΔT.*

Fig.3b shows the tunnel current density in parallel magnetic configuration versus difference of voltage between the two electrodes and difference of temperature. For $\Delta T=0$, the tunnel magnetoresistance (TMR) of the MTJ could be calculated from the current in parallel and antiparallel configurations: TMR=9.2% for the used parameters.

In the media in operation, no net charge current flows across the grain boundaries: J(V, $\Delta T$)=0. This means that the effect of the difference of temperature between neighboring grains is counterbalanced by the onset of a thermoelectric voltage between grains. The hotter grain tends to emit more electrons than the colder grain which is compensated by the onset of a positive voltage on the hotter grain with respect to the colder one. The line J(V, $\Delta T$)=0 representing the thermoelectric voltage versus $\Delta T$, $V_{th}(\Delta T)$ is extracted from Fig.3b and plotted in Fig.3c. With a temperature dependence of 60K between neighboring cooling grains in HAMR media, $V_{th}$ can reach about 1mV which is quite large.

Next the two components of the torques $T_{//}$ and $T_{\perp}$ were calculated first as a function of the two parameters V and $\Delta T$ Fig.4a and c) and then by considering the condition J=0 i.e. V= $V_{th}(\Delta T)$ (Fig.4b and d).

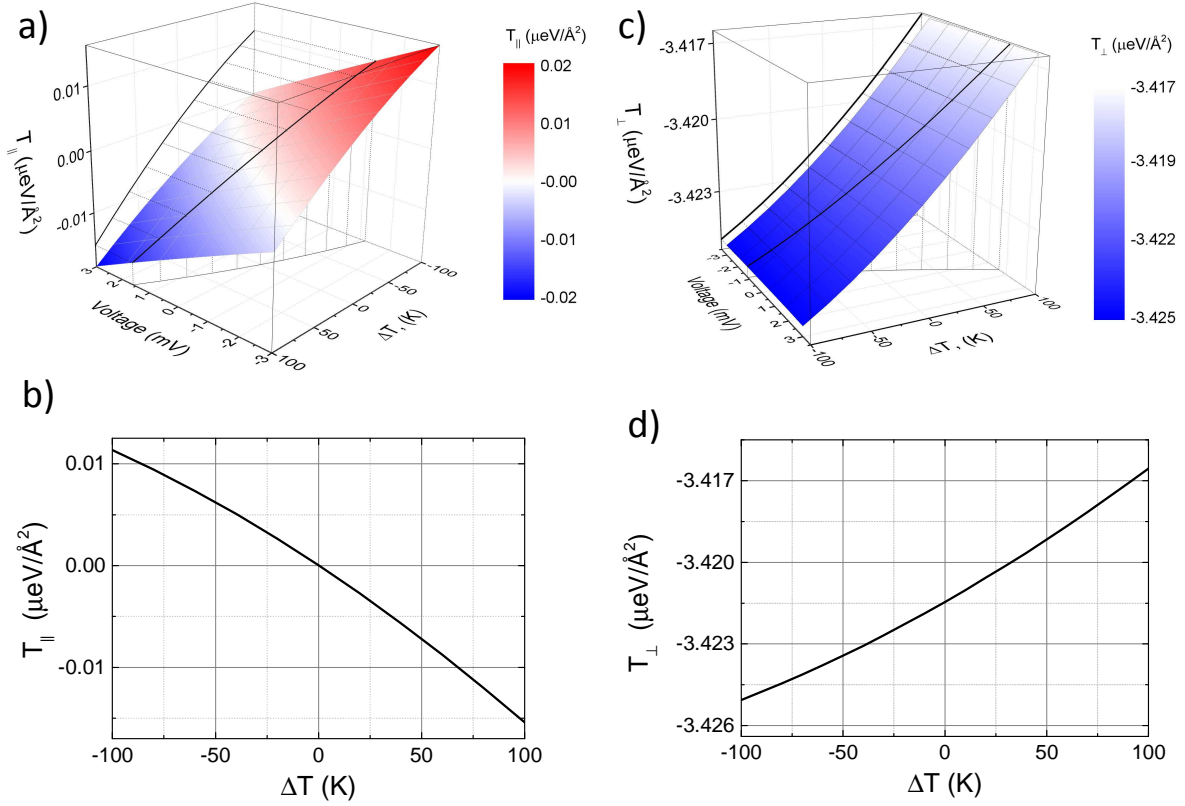

*Figure 4: a) In-plane component of the spin-transfer torque as a function of bias voltage and difference of temperature between the two electrodes. The line in the (V, ΔT) plane represents the thermoelectric voltage $V_{th}$ versus ΔT as plotted in Fig.3c. The line in the ($T_{//}$, ΔT) plane represents the in-plane torque due to the thermal gradient across the tunnel barrier for $V=V_{th}(ΔT)$. This line is reported in Fig4b. b) In plane component of the thermal spin-transfer torque versus the difference of temperature between the two electrodes. c) Same as Fig.a for the perpendicular component of the torque. Note that the vertical scale is here much more expended than for the in-plane case. d) Same as Fig.b for the perpendicular component of the thermal spin-transfer torque.*

For a difference of temperature between neighboring grains of typically 60K, in-plane thermal torques of the order of -0.01 μeV/Å² can be expected (Fig.4a). Such torques can have a quite significant impact during the cooling of the grains below Tc considering that the thermal stability factor Δ of these grains just start to increase from Δ=0 at Tc to Δ~10 at Tc-10K. For comparison, in MgO based magnetic tunnel junctions for spin transfer torque magnetic random access memory (STT-MRAM) which exhibit much higher TMR and have a thermal stability factor Δ=80, in-plane torques of the order of 0.16 μeV/Å² [43] are sufficient to induce switching of the storage layer magnetization. Since the torque required to switch the magnetization scales with Δ, this means that in the present context of HAMR, the predicted torque amplitude of 0.01 can switch the magnetization of the grains when their thermal stability factor is still below ~5. As a result, a significant influence of this in-plane TST can be expected on the grain magnetization dynamics in the initial stage of their cooling down below T. This will be further confirmed by the micromagnetic simulations. Concerning the out-of-plane component (Fig.4d), the amplitude of its variation as a function of ΔT is about 3 times weaker than the in-plane component but it is centered around a large value at ΔT=0. As for the perpendicular STT in MTJs, this out-of-plane component of the torque at ΔV=0 or ΔT=0 corresponds to an exchange like coupling across the barrier. This coupling is often balanced by other energy terms. For instance in STT-MRAM, the influence of this coupling across the barrier can be balanced by the stray field from the reference layer. In the context of HAMR media, it can be balanced by the dipolar coupling between neighboring grains. In any case, as we will see in the simulation part, this perpendicular torque may influence the precessional frequency during the grain magnetization switching but does not play a significant role on the magnetization switching probability itself.

Next, the in-plane and perpendicular TST amplitudes per K of temperature difference between the two neighboring grains were calculated. By analogy with the spin transfer torkance whose respective in-plane and out-of-plane components are $\frac{dT_{//}}{dV}$ and $\frac{dT_{\perp}}{dV}$ [44], one can define a thermal torkance having two components $\frac{dT_{//}}{d\Delta T}$ and $\frac{dT_{\perp}}{d\Delta T}$. These two components are plotted in Fig.5 versus the tunnel barrier RA product. For the chosen value of RA=0.3Ω.μm², the in-plane thermal torkance is about 3 times larger than the out-of-plane one and the two have opposite signs.

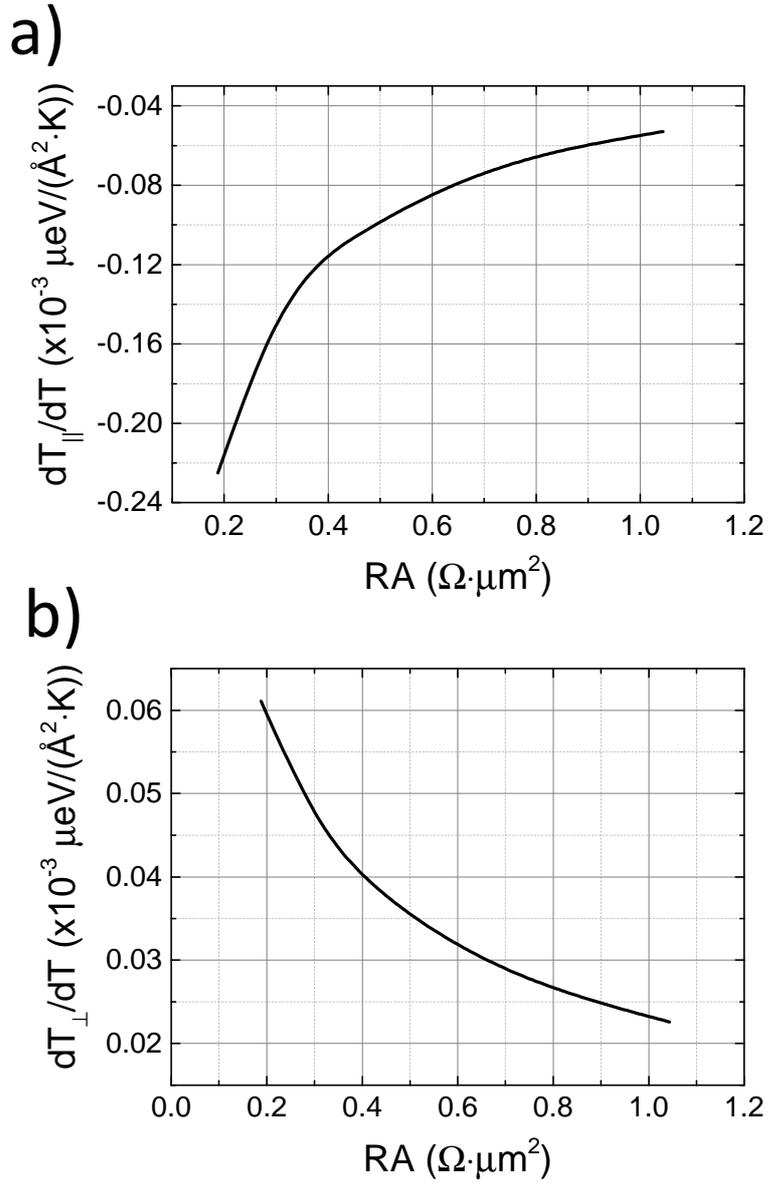

*Fig.5: a) In-plane thermal torkance; b) Out-of-plane thermal torkance, versus Resistance x Area product (RA).*

In the above calculations, we used reasonable values for the parameters representing FePtC HAMR media, in particular for the Stearns polarization of FePt and for the C barrier height. We checked the relevance of these parameters in actual HAMR media by performing electrical measurements on real media as presented in the next section.

**Determination of the spin polarization of electrons tunneling across grain boundaries in FePtC HAMR media**

As seen in the previous section, key parameters which determine the TST amplitude for a given temperature difference between neighboring grains are the grain boundary RA product and the spin polarization of the tunneling electrons. To determine these parameters, FePtC HAMR media were deposited and characterized from magnetic and transport points of view.

FePt-35.5 vol.%C 10 nm films were prepared by DC magnetron sputtered on single-crystalline MgO (001) substrates. The magnetic film was directly deposited on MgO substrate to avoid any electrical shunting in the heat sink, SUL or any conducting underlayer material during the electrical characterization. Details on the film preparation are given in the method section.

The structure and the degree of L10 order were characterized by X-ray diffraction (XRD) and the microstructure of the films by transmission electron microscopy (TEM) (Fig.2b). The grain size was found to be ~8nm (Fig.2c).

The films were magnetically characterized at room temperature with an applied magnetic field of up to ± 7 T. A clear perpendicular anisotropy was seen with an out-of-plane coercivity of 3.68T (See Fig.6). The in-plane loop exhibits a very slight remanence which indicates that a small fraction of grains may be misoriented and exhibit an in-plane axis of anisotropy as often observed in FePt based HAMR media[29].

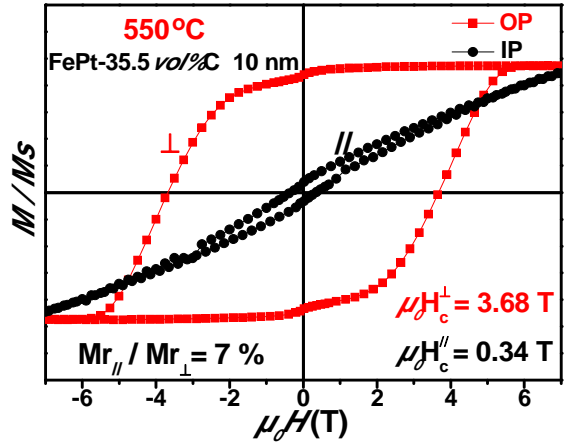

*Figure 6: In-plane and out-of-plane hysteresis loops measured at 300K on the FePtC 10nm film shown in Fig.2b and 2c.*

For the electrical measurements, the FePtC nano-granular thin film was patterned in the form of a strip with current and voltage contacts by electron-beam lithography (inset of Fig.7a). Fig.7a and b respectively show the V(I) characteristics and the magnetoresistance of the sample between ± 14 T measured at 300K.

The V(I) characteristic is linear in the range of current investigated, which is not surprising since the sample can be viewed as a set of ~1.4 $10^5$ MTJs connected in series (see section method) so that the voltage across each junction is very small. Since the inner part of the grains is metallic while the grain boundaries are insulating or semiconducting, one can assume that all the voltage drop is across the grain boundaries or in other words that the measured resistance originates from all the grain boundaries connected in series. Knowing the grain size and therefore the number of grain boundaries traversed by the current along its path between the two voltage contacts, the grain boundary RA can be determined (see section method): RA=0.28 $\Omega.\mu m^2$, very close to our chosen theoretical value of 0.30 $\Omega.\mu m^2$.

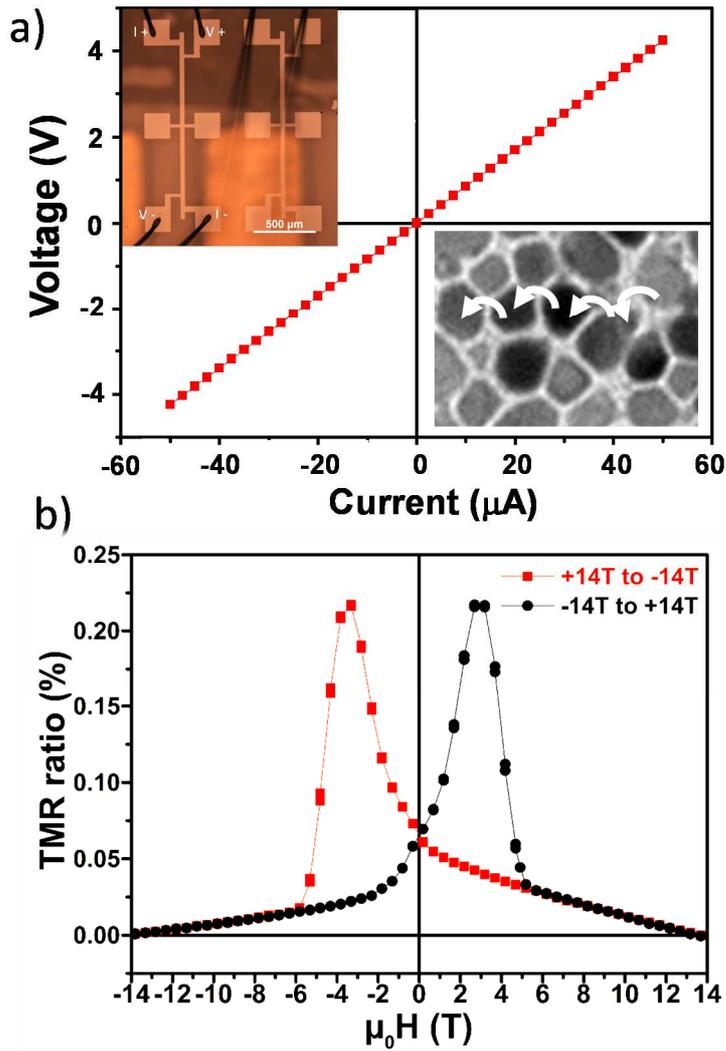

*Figure 7: a) Current in plane V(I) characteristic and b) magnetoresistance loop measured at 300K with field out-of-plane. The Upper left inset of Fig a shows the patterned sample stripe with electrical contacts. The bottom right inset is a qualitative representation of electrons tunneling from grain to grain across the insulating (semiconductor) grain boundaries.*

Concerning the magnetoresistance, it has a double peak shape with maxima at the same fields as the coercive fields measured on the hysteresis loops (Fig.6). This magnetoresistance corresponds to a giant magnetoresistance (GMR) or rather a tunnel magnetoresistance (TMR) signal since the grain boundaries are insulating or semiconducting. This TMR signal can be understood as follows: At positive and negative saturations, all grains are in parallel magnetic configuration yielding a low resistance state. Starting from positive saturation and gradually reducing the field, thermally activated grains magnetization fluctuations increase yielding a slight increase of resistance down to zero field. Then by applying a negative field of increasing amplitude, as the field approaches the coercive field, grains start switching in the negative direction. This locally creates antiparallel alignment between neighboring grains which increases the system resistance. The proportion of antiparallel grains is maximum at the coercive field yielding the maximum of resistance. Beyond Hc, the resistance associated with the TMR contribution decreases again and becomes minimum when the grains reach negative saturation. The measured amplitude of magnetoresistance is relatively weak: 0.23%. However, one has to consider that this magnetoresistance is influenced by the proportion of neighboring grains which are in antiparallel alignment. Indeed, if at the coercive field, the media consists of alternating up and down domains each one containing N grains of parallel magnetization along the current path, this means that there is only a fraction $1/N$ of junctions which are in antiparallel alignment. Consequently, the measured TMR has only $1/N$ of the value that it would have if all the junctions were in antiparallel alignment.

In order to estimate *N*, magnetic force microscopy (MFM) experiments were carried out on this FePtC HAMR media, with an expected spatial resolution around 25nm based on the tip and imaging parameters (see section method for more details). A typical image of the media at the negative coercive field after positive saturation is shown in Fig.8.

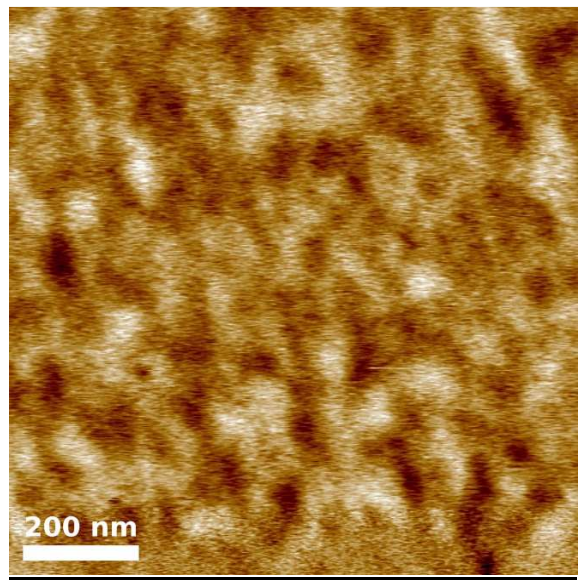

*Figure 8: Magnetic Force Microscopy image with lift height 10nm and tip oscillation amplitude 16nm. The peak-to-peak phase contrast is 0.4°.*

Clearer and darker contrasts are observed from which a domain length scale of the order of 80nm can be derived. Considering that the grain size is ~8nm, this implies $N$~10. Therefore the actual TMR of each individual junction between parallel and antiparallel alignment can be considered to be ~2.3%. This is about 4 times lower than our expected theoretical value of 9.2% reported above. The difference can be due to several factors: i) the calculated value is at 0K whereas the experimental one is at 300K; ii) the ab-initio calculation from which the parameters were chosen for FePt did not take into account spin orbit effects which may reduce the polarization of FePt; iii) experimentally, some spin flip impurities (Pt) may diffuse to the grain boundaries further reducing the polarization of the tunneling electrons. According to Julliere's expression of the TMR, TMR=$2P^2/(1-P^2)$[45], the TMR is roughly proportional to $P^2$ for low polarization values. Consequently, this 4x reduction in the TMR amplitude means that the polarization of the tunneling electrons in our actual HAMR media may be about half of that estimated from ab-initio calculations. Correlatively, the actual TST torques in the HAMR media may be half of the theoretical expected value, but the orders of magnitude remain the same.

We next investigated the impact of these TST torques on the written patterns in HAMR as discussed in the following section.

**Micromagnetic simulations of HAMR writing taking into account the thermal spin transfer torques**

In the simulation, the FePtC granular recording media was represented by an assembly of Voronoi cells[46]. The averaged FePt grain size was set to 8.0nn and their thickness to 8.5nm. Non-magnetic boundary regions representing the grain boundaries were also created by shifting the vertices of each Voronoi cell by 0.5nm towards the seed point of each Voronoi cell. A thermal gradient of 6.6K/nm was assumed between neighboring grains which is a typical value in HAMR recording (see Fig.1c). The media was assumed to move at a linear velocity of 22m/sec corresponding to a rotating speed of 10kRPM at a radius of 21mm. In-plane $T_{//}$ and out-of-plane $T_\perp$ TST terms were added to the LLG equation to calculate the writing dynamics combined with the thermal dynamics (Eq 1, 2, 3, in GGS)[47]. γ and α are respectively the gyromagnetic parameter and Gilbert damping. $M_1$ and $M_2$ are the magnetization vectors of neighboring grains with saturation magnetization $M_s$. $S_{boundary}$ is the contact boundary area between neighboring grains, equal to 48.8 nm² in average for this model. Considering that some uncertainty exists on the actual amplitudes of $T_{//}$ and $T_{\perp}$ as discussed above, these TST amplitudes were varied from 0 to 100 % of their calculated expected values (Fig.4b and d) for FePtC media to evaluate their impact on the HAMR writing process.

$$\frac{d\vec{M_1}}{dt} = -\gamma \left\{ (\vec{M_1} \times \vec{H}) + \frac{\alpha}{Ms} \cdot \vec{M_1} \times (\vec{M_1} \times \vec{H}) + S_{boundary} \cdot (\vec{\Gamma_{//}} + \vec{\Gamma_{\perp}}) \right\} \quad (1)$$

$$\vec{\Gamma_{//}} = T_{//} \cdot \vec{M_1} \times (M_1 \times \vec{M_2}) \quad (2)$$

$$\vec{\Gamma_{\perp}} = T_{\perp} \cdot (\vec{M_1} \times \vec{M_2}) \quad (3)$$

In the simulations, the bit length was set to 40nm and track width to 80nm. This corresponds to a recording density of 635kFCI x 317kTPI = 202Gbit/in² which is relatively low to better evidence the disturbance induced by the TST torques on the magnetic orientation of the grains. During write, a 1T up or down vertical magnetic field was applied on the media in the perpendicular direction.

Figure 9 shows the impact of the thermal gradients on the bit patterned for various amplitudes of the TST torques ranging from 0 to 100% of the calculated value expected for FePtC as discussed above. Clearly, the impact is quite significant as soon as the torque becomes of the order of 50% or more of the expected theoretical value (Fig10c-e). Qualitatively this impact can be understood as follows: As the grains in the media move away from underneath the plasmonic antenna, they cool down very quickly in the magnetic field generated by the writer. Let us assume a thermal gradient of 10K/nm with a grain size of 8nm and consider two neighboring grains in the downtrack direction, one just crossing the line $T=T_C$ while the neighboring grain is already at Tc-80K. At this temperature, the cooler grain has already gained about 50% of the value of its magnetization at 300K. Its anisotropy field already exceeds several Teslas[3] so that the cooler grain can be considered already as pinned. In contrast, the hot grain just starts building up its magnetization. Due to its high temperature, it emits by thermoemission up and down spin polarized electrons. However mostly electrons with spin parallel to the magnetization of the cooler grain are emitted as in magnetic tunnel junctions[39]. Consequently, a spin accumulation opposite to the magnetization of the cool grain builds up in the hot grain favoring antiparallel alignment between the cool grain and the hot grain, similarly to what was observed in MTJs[39]. If the grains were all well aligned this would favor antiparallel alignment between grains in the downtrack direction which would actually improve the sharpness of the transition while degrading the bulk part of the bits. However, due to the misalignment of the grains, the net outcome looks more like a random noise which drastically increases with the TST amplitude.

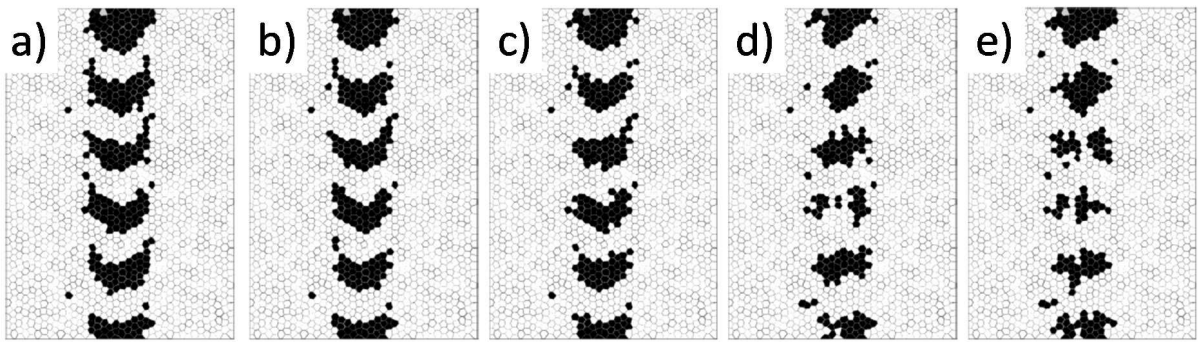

*Figure 9: Simulated images (300nm x 440nm) of tracks of alternating 0 and 1 bits assuming various amplitudes of the thermal torques ranging from 0 to100% of the calculated torques expected for FePtC media. a) No TST; b) 25% TST c) $T_{//}$ = 50% calculated TST ; d) 75% TST ; e) 100% TST*

Finally, the relative influence of the in-plane and perpendicular torques on the bit patterns was also investigated (Fig.10). Different situations were compared with both torques being present (Fig.10a similar to Fig.9d) or only one or the other of the torque components being taken into account (Fig11b: only perpendicular torque ($T_{//}$ =0), Fig11c: only in-plane torque ($T_\perp$ =0)).

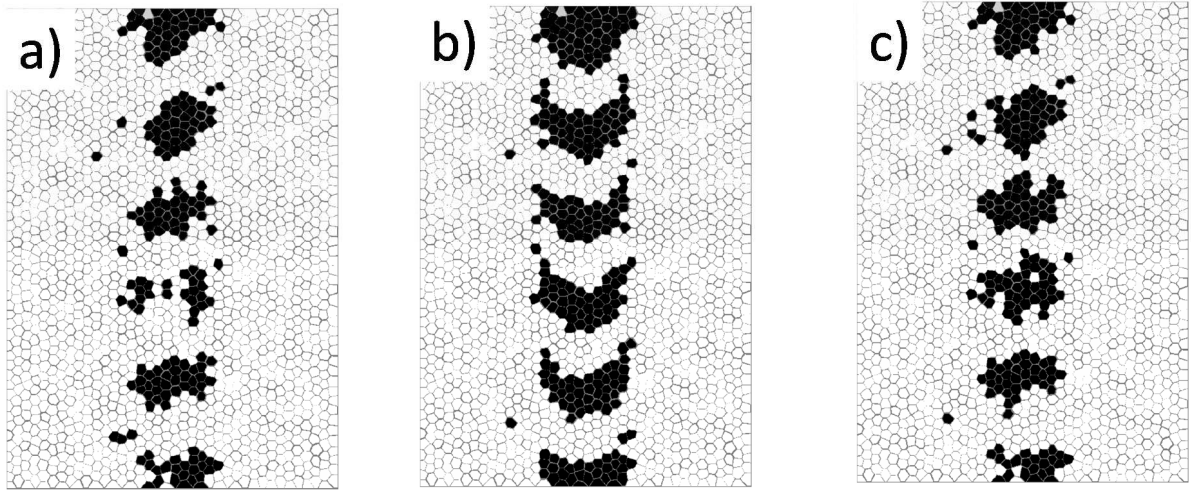

*Figure 11: Images similar to Fig.10 in which the relative influence of the two torque terms was varied: a) Both $T_{//}$ and $T_\perp$ are taken into account (as in Fig.10d) ; b) Only $T_\perp$ is taken into account; c) Only $T_{//}$ is taken into account.*

Clearly, mostly the in-plane thermal torque impacts the recording quality while the perpendicular thermal torque has a much weaker impact. This is quite consistent with earlier observations on out-of-plane magnetic tunnel junctions showing that the magnetization switching is mostly influenced by the in-plane spin transfer torque (STT) and not so much by the perpendicular STT[48].

**Discussion and Conclusion**

This study is pointing out the importance of a new phenomenon taking place during HAMR writing: thermal spin transfer torque between grains which tends to favor antiparallel alignment between the already cold grains and those which are just starting to cool down below the Curie temperature. Our theoretical and experimental results demonstrate that the effect can be quite significant in HAMR media and has a negative impact on the recording performances. If the thermal gradient is further increased as required according to HAMR technology roadmap[29] (up to 14K/nm for 2Tbit/in² and 18K/nm for 4Tbit/in²), the TST amplitudes would be tripled compared to the present estimate. According to Fig.9, this should have a very strong impact on recording patterns. In principle, if the grains were perfectly aligned, the effect could improve the transition but degrade the inner part of the bits. However, since the grains are not perfectly aligned with respect to the line of Tc crossing and because of the grain to grain Tc distribution, the TST yields a random media noise caused by randomly distributed magnetically misoriented grains. In terms of media design, this implies that the effect should be minimized as much as possible. This means reducing the thermal spin current which flows between grain boundaries during the media cooling. This can be done by increasing the resistance of the grain boundaries (see Fig.5a). C was chosen because it yields a better texture of the FePt alloy and improves the cylindrical shape of the grains. However, being a semiconductor material, it is relatively transparent to the electrons. Oxide barriers can provide higher barrier heights (SiOx, AlOx, MgO…) as well as nitrides but the media texture is not as good as with C. Another possibility is to try to reduce the spin polarization of the spin current by introducing spin flip impurities in the grain boundaries. These should be heavy metal impurities (Ta, W, Mo, Bi etc) which would segregate to the grain boundaries together with the host segregant. One may also play with the underlayer or capping layer of the magnetic grains by introducing heavy metal layer or impurities acting as spin sink to evacuate the spin accumulation which builds up in the cooling grains.

For experimentalists working in spincaloritronics, our study also points out the huge thermal gradients which can be produced by plasmonic antenna. The gradients are one to two orders of magnitude larger than the gradients used so far in spincaloritronics experiments. The use of such extremely high gradient may yield new effects or dramatically increase the amplitude of previously observed effects. One should however notice that the huge thermal gradients in HAMR are obtained thanks to a combination of tiny heat source (the plasmonic antenna, heating area~50nm) and media motion (20m/s) which allows ultrafast cooling of the media. Due to heat diffusion, in experiments without any moving parts, one should work in pulsed mode (for instance ns laser pulse or faster) to allow ultrafast cooling following each pulse and therefore reach such large gradients.

## **Methods:**

FePtC media deposition and characterization:

FePt-35.5 vol.%C 10 nm film was DC magnetron sputtered on single-crystalline MgO (001) substrates. Firstly, the substrate was thermally cleaned at 600°C for 1 h, then at the same substrate temperature a FePt-C granular layer was deposited by co-sputtering Fe, Pt and C targets under an Ar pressure of 0.48 Pa with a deposition rate of about 0.16 nm/s. Thereafter, the film was vacuum cooled to room temperature and a 3 nm carbon over coating layer was deposited on top of FePt-C granular layer as protection layer. The thickness of the film and the atomic/volume fraction of Fe, Pt, and C were estimated using the pre-calibrated sputtering rates. The chemical composition of FePt was estimated to be nearly 1:1 from inductively coupled plasma (ICP) analysis. An alternating layer deposition method was used for FePt-C film growth to suppress the growth of randomly oriented spherical grains on the (001) textured FePt granular layer[49], and the FePt-C film thickness was optimized to ensure a single layered structure[50].

The structure and the degree of L10 order were examined using X-ray diffraction (XRD) with Cu Kα radiation ($\lambda$= 1.542 Å). The microstructure of the films was characterized by transmission electron microscopy (TEM) using FEI Tecnai T20. The room temperature magnetization curves were measured by superconducting quantum interference device vibrating sample magnetometer (SQUID-VSM) with an applied magnetic field of up to ± 7 T.

For the magnetoresistance measurement, the FePtC nano-granular thin film was patterned by electron-beam lithography in the form of an elongated strip with current and voltage contacts (inset Fig.7a). The electrical transport properties were investigated using a standard dc four point probe system with a constant current source ranging from 0.1 µA to 100 µA. The measurements were performed using a 14T Quantum Design Dynacool physical property measurement system (PPMS). The resistance R of a sample L=1100µm long, w=45µm wide and t=10nm thick was measured to be 84950Ω. Assuming a grain size ~8nm, the current traverses 137500 grain boundaries. The RA of each boundary is then given by RA=R*w*t/137500=0.28Ω.µm² very close to our chosen theoretical value of 0.3 Ω.µm².

Magnetic Force Microscopy (MFM) characterization:

Magnetic Force Microscopy measurements are performed in air with an NT-MDT NTegra instrument. Probes are commercial AC240TS cantilevers capped by a 10nm-thick sputter-deposited layer of a CoCrPt alloy. Measurements are performed in a two-pass scheme, the topography being monitored in the first pass with an amplitude modulation feedback, while the magnetic signal is monitored in the second pass through the phase. A voltage is applied between tip and sample, and adjusted to minimize electrostatic contributions and thus cross-talk with topography. The lift height is 10nm and the tip oscillation amplitude is 16nm, which for the tip coating used is expected to yield a spatial lateral resolution around 25nm. Thus, grains may not be resolved individually.

**Acknowledgements**

Kouji Shimazawa and Moris Dovek from TDK/Headway are warmly acknowledged for fruitful and stimulating discussions.

This work was partially supported by the European Commission through ERC Adv grant MAGICAL n°669204.

**Authors contributions**

Bernard Dieny has initially thought about this potential impact of thermal spin transfer torque in HAMR recording performance and conceive the study. He coordinated the collaboration.

Mair Chshiev has developed the theory and code to calculate the transport properties across magnetic tunnel junctions submitted to bias voltage and thermal gradient in free electron model.

Brian Charles and Nikita Strelkov have performed the theoretical calculations of the thermal torques in various situations close to FePtC media.

Alain Truong and Olivier Fruchart have performed magnetic characterization (MFM) of the HAMR media.

Ali Hallal has performed the ab-initio calculation on FePt (supplemental material) to determine the band splitting to be used in the free electron model.

Jian Wang, Yukiko K. Takahashi, Kazuhiro Hono frol NIMS have grown the FePtC media, measured their magnetic hysteresis loops, patterned the films for electrical measurements and measured their current in plane magnetoresistance.

Tomohito Mizuno from TDK/Headway Technologies has performed the micromagnetic simulations.

**Competing financial interests:**

The authors declare having no competing financial interest.

# Impact of intergrain spin transfer torques due to huge thermal gradients on the performance of heat assisted magnetic recording

Bernard Dieny[1a], Mair Chshiev[1], Brian Charles[1], Nikita Strelkov[1], Alain Truong[1], Olivier Fruchart[1], Ali Hallal[1], Jian Wang[2], Yukiko K. Takahashi[2], Tomohito Mizuno[3], Kazuhiro Hono[2]

**SUPPLEMENTAL MATERIAL**

### A) Ab intio calculation of FePt electronic band structures

Our first principles calculations were performed using the Vienna ab-initio simulation package (VASP) [1-2] with generalized gradient approximation [3] and projector augmented wave potentials [4-5]. A 19×19×13 K-point mesh was used in the calculations with the energy cut-off equal to 520 eV. Initially the structure was relaxed in volume and shape until the force acting on each atom falls below 1 meV/Å. The FePt structure is a body centered tetragonal with lattice parameter a=b=2.73Å and c=3.75Å [c.f Figure1(a)]. The Kohn-Sham equations are then solved to find the charge distribution of the ground state system. Next the band structure was calculated along the high symmetry K-points ΓX and ΓZ as shown Figure1 (c) (d). The red and blue solid lines represent the spin-up and spin-down, respectively. By focusing on the parabolic bands (exchange splitted free-like electrons as described by Stearns in Ref 41 of the manuscript), the exchange splitting around Fermi level was found to be around 2.1-2.5 eV along the high symmetry points ΓX and ΓZ (see Figure 2a and b). Since we found all calculated splitting values between 2.11-2.5 eV, we chose 2Δ=2.2eV for our calculation of thermal spin transfer torque.

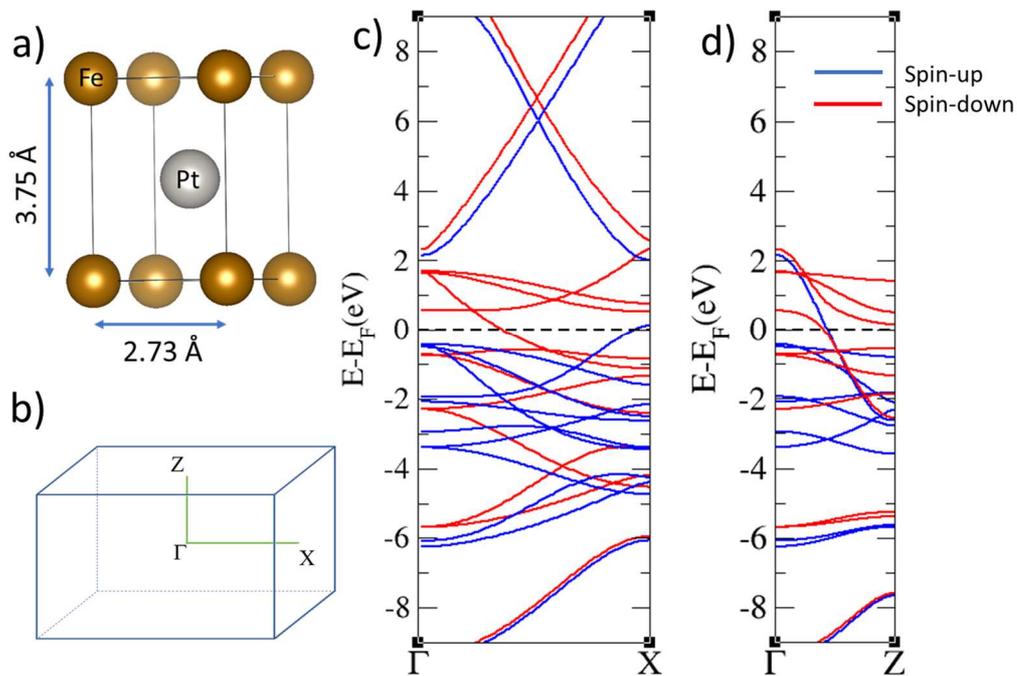

*Figure **Erreur ! Document principal seulement.** a) $L_{10}$ FePt body centered structure. b) High symmetry K-points in Brillouin zone used for band structure calculations. c) and d) Spin-dependent electronic structure of FePt along ΓX and ΓZ, respectively. Spin-up and spin-down bands are presented by blue and red, respectively.*

a)

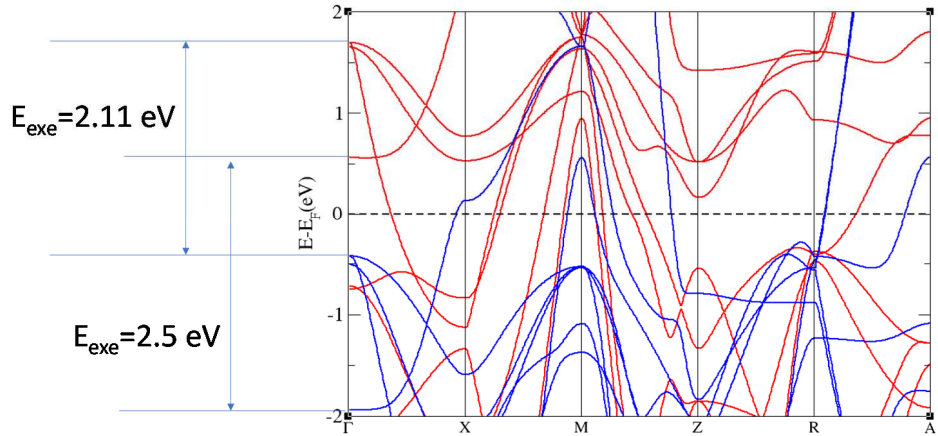

b)

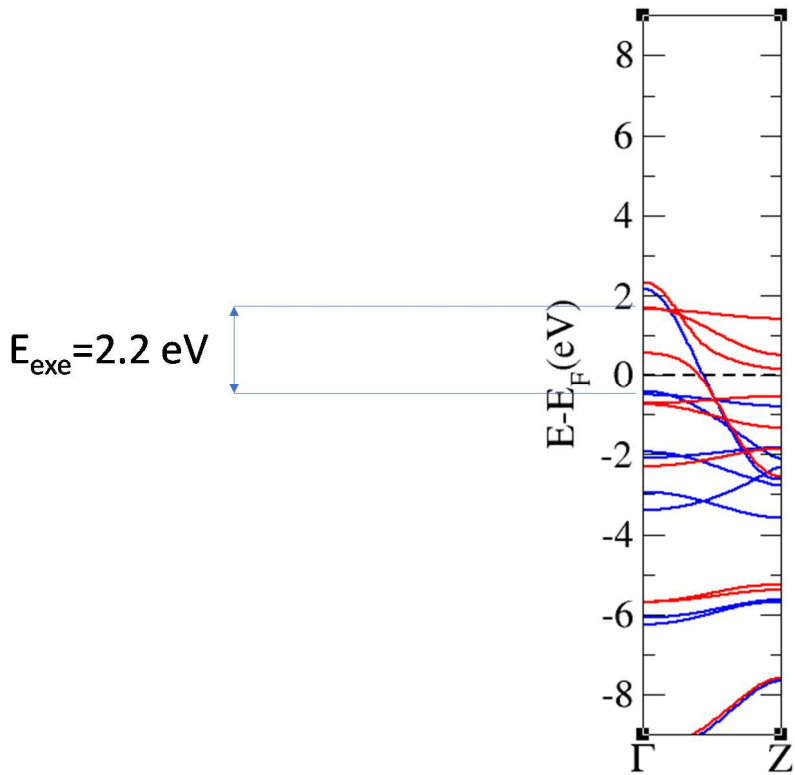

*Figure 2 Examples of determination of the band splitting parameter used 2Δ in our theoretical approach (used value 2.2eV) to calculate the thermal spin transfer torques in the model of band splitted free electrons as first proposed by Stearns.*

## B) Estimation of the in-plane STT torque to switch magnetization in conventional STT-MRAM

According to equation (21) in Ref.1, the in-plane spin transfer torque is given by $T = -\tau_L \frac{\hbar}{2e} J_0 \, \boldsymbol{l} \times (\boldsymbol{r} \times \boldsymbol{l})$ where $\tau_L$ is the spin polarization (~80% in MgO based MTJ), $\hbar$ is the reduced Planck constant, $e$ the electron charge, $J_0$ the current density. The coefficient $J_0 \hbar/(2e)$ is in units J/m². In state of the art MTJs for spin transfer torque MRAM, the switching current density are of the order of $J_0 = 10^6$ A/cm². The torque in $\mu eV/A^2$ can then be calculated as follows:

$$T_{//} = \tau_L \frac{\hbar}{2e} J_0 = 0.8 \frac{1.05 \cdot 10^{-34}}{2 \cdot 1.6 \cdot 10^{-19}} \, 1 \cdot 10^{10} = 2.63 \cdot 10^{-6} \frac{J}{m^2}$$

$$2.63 \cdot 10^{-6} \frac{J}{m^2} \times \frac{10^{-20}}{1.6 \cdot 10^{-19}} 10^6 = 0.16 \frac{\mu eV}{A^2}$$

Reference:

[1] J.Slonczewski, PRB 71, 024411 (see equation 21).